\documentclass[11pt]{article}
\usepackage{moriond,epsfig}

\def\Journal#1#2#3#4{{#1} {\bf #2}, #3 (#4)}

\def\be{\begin{equation}}
\def\ee{\end{equation}}
\def\bea{\begin{eqnarray}}
\def\eea{\end{eqnarray}}

\begin{document}
\vspace*{4cm}
\title{ANTARES SENSITIVITY TO DIFFUSE HIGH ENERGY NEUTRINO FLUXES}

\author{ ALAIN ROMEYER }

\address{DAPNIA/DSM/SPP, CEA Saclay \\
91191 Gif sur Yvette CEDEX, France}

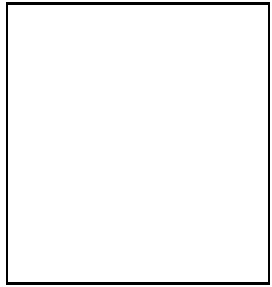
\maketitle
\abstracts{
The European collaboration ANTARES aims at operating a large deep-sea
neutrino telescope in the Mediterranean sea. The detection of high-energy 
cosmic neutrino can improve our knowledge on the most powerful astrophysical 
sources in the Universe and about the origins of cosmic rays. A first 
Monte-Carlo study for the ANTARES sensitivity to diffuse neutrino fluxes 
predicted by current models is reported.}

The goal of the ANTARES project is to observe Cherenkov light produced 
by high-energy muons induced by neutrino interactions in the matter 
surrounding the detector\cite{proposal}. The neutrinos to be detected 
can most likely originate from : cosmic-accelerators, cosmic ray showers 
in the atmosphere or neutralino annihilations. Moreover, neutrinos are 
the only particles with all the required properties for high energy 
astronomy : they are neutral (no deflection by magnetic fields), stable 
and interact weakly (travel through cosmological distances and escape 
from dense regions of the universe). At high energy, the muon trajectory 
is closely aligned with the incident neutrino, so it indicates 
the source position in the sky with a good angular resolution of 0.2$^o$ at 
high energy. ANTARES can therefore probe the fundamental physics of cosmic 
ray origins, neutrino oscillations and dark matter in the Universe.\\

The detector will be installed in the Mediterranean sea, at a depth of 
2400 m, 40 km from La Seyne sur Mer (near Toulon, France). It will 
consists of 10 identical strings, each about 400 m high, supporting 30 
storeys. Each storey has 3 downward looking optical modules. Each optical 
module is essentially a pressure resistant glass sphere containing a 10 inch 
photomultiplier (PMT). The PMTs detect the times and amplitudes for hits 
of Cherenkov light emitted by the muons when crossing the sea water. This 
information allows the reconstruction of the muon track with a good 
precision. The PMT positions will be determined to the required accuracy 
($\sim$10 cm) by an acoustic positioning system composed of hydrophones 
and emitters and by a system of tiltmeters and compasses along the 
strings. 
 
The astrophysical mechanisms likely to produce high energy neutrinos in 
general, are hadronic processes involving pion decay. Potential sources 
are particle accelerators in the Universe : extra-galactic sources like 
Active Galactic Nuclei (AGN) and Gamma Ray Bursts (GRB) or galactic sources 
like supernova remnants and binary stars. 
	
The background comes from the shower development initiated by the interaction 
of a cosmic ray in the atmosphere. Atmospheric muons and neutrinos are 
produced by meson decays. The muon flux, which is the dominant component 
of the background, is suppressed by rejecting downward-going tracks. The 
atmospheric neutrino flux is composed of  : the conventional neutrino flux, 
due to pion and kaon decays, dominant at low energy,  known with an 
uncertainty of about 20 $\%$ ; and prompt neutrinos, due to the decay 
of heavy flavour particles, can be dominant at high energy. The discrepancy 
between different prompt neutrino flux predictions can reach two orders of 
magnitudes \cite{costa}.

Diffuse neutrino fluxes based on unresolved source distributions are 
predicted by various models\cite{P96,SDSS91,WB98} leading to an event rate 
between 1 to 100 events per year in the ANTARES detector. The signal can be 
identified as an excess above the atmospheric neutrino background of high 
energy events. The muon energy estimator is based on the amount of light 
detected and leads to an energy measurement within a factor 3 of the true 
value above 1 TeV \cite{nrj}. Monte-Carlo simulations show that the signal 
exceeds the background for energies above 1-10 TeV as illustrated in the 
Table \ref{tab:nevt}.
\begin{table}[h]
\caption{Integrated number of events per year above a given Monte-Carlo 
muon energy threshold.
\label{tab:nevt}
}
\vspace{0.4cm}
\begin{center}
\begin{tabular}{|l|c|c|c|} \hline
&\multicolumn{3}{c|}{Energy threshold}\\
& 1 TeV & 10 TeV & 100 TeV \\ \hline
atmospheric (conv.) & 240 & 19 & 0.2 \\ \hline			
atmospheric (conv. + prompt) & 243 - 273 & 20 - 30 & 0.2 - 1.1 \\ \hline
P96 \protect\cite{P96} & 29 & 25 & 14 \\ \hline
SDSS91 \protect\cite{SDSS91} & 136 & 129 & 76 \\ \hline
\end{tabular}
\end{center}
\end{table}
As an example, two different AGN models of diffuse neutrino fluxes have been
considered : one by Protheroe \cite{P96} referred to as P96 where 
protons interact with radiations produced by the accretion disk, and another 
one by Stecker et al.\cite{SDSS91} referred to as SDSS91 based on X-ray 
observations (already ruled out by the experimental upper 
limit given by the AMANDA collaboration\cite{AMANDA}). Table \ref{tab:nevt} 
shows the expected number of events per year in the detector : above 100 TeV 
this number is between 0.2 and 1.1 for the atmospheric neutrino background, 
depending on the prompt neutrino flux used, around 14 (76) for 
the signal using the P96 (SDSS91) model.\\

The construction of the ANTARES detector is under way. Deployment of complete 
strings will start in 2003 and are due to terminate in 2005. After a few 
years of data taking, the ANTARES experiment will be able to measure the 
cosmic neutrino fluxes and therefore add new information on possible sources 
of high energy cosmic neutrinos.

\end{document}